\documentclass[%
 reprint,
 amsmath,amssymb,
 aps,prc,
]{revtex4-2}
\usepackage{braket}
\usepackage{graphicx}
\usepackage{dcolumn}
\usepackage{bm}
\usepackage{hyperref}
\usepackage{natbib}
\usepackage{amsmath}
\usepackage{booktabs}
\newcommand{\RNum}[1]{\uppercase\expandafter{\romannumeral #1\relax}}
\setcitestyle{bibnames=5}

\begin{document}


\title{Unraveling the Effects of Cluster Transfer-Induced Breakups on $^{12}$C Fragmentation in Hadron Therapy}
\author{Arunima Dev T V}
\author{Anagha P. K}
\author{Midhun C.V}\thanks{midhun.chemana@gmail.com}%
\author{M.M Musthafa}\thanks{mmm@uoc.ac.in}%
\author{Vafiya Thaslim T.T}
\author{Nicemon Thomas}
\author{Antony Joseph}
\affiliation{Dept. of Physics,University of Calicut, Calicut University P.O Kerala, 673635 India}

\author{S. Ganesan}
\affiliation{Formerly Raja Rammanna Fellow of the DAE, Bhabha Atomic Research Center, Mumbai, 400085, India}

\date{\today}

\begin{abstract}
The capability of standard Geant4 PhysicsLists to address the fragmentation of $^{12}$C$-^{12}$C was assessed through a comparative analysis with experimental cross sections reported by Divay et al.\citep{Divay} and Dudouet et al.\citep{Dudouet}. The standard PhysicsLists were found to be inadequate in explaining the fragmentation systematics. To address this limitation, the breakup component of fragmentation was systematically integrated into the standard PhysicsList, which successfully replicated the differential and double differential cross sections for $\alpha$ production. This breakup component was modeled using {\sc fresco} CDCC-CRC calculations. This novel physics process was then incorporated into the Geant4 framework, facilitating the calculation of dose distributions in water and tissue. The application of this method demonstrated a precise reproduction of the dose deposited at the Bragg peak region, corroborating the experimental data from Liedner et al.\citep{Leidner_2018}, thereby enhancing the accurate visibility of dose tailing.

\end{abstract}
\maketitle
\section{Introduction}
Carbon therapy is recognized as a promising modality in radiotherapy, owing to its advantageous physical and radiobiological characteristics \citep{Amaldi_2005}. Nevertheless, concerns persist regarding the generation of secondary charged particles stemming from nuclear reactions with isotopes within the tissue \citep{Fragmentation}. These secondary particles not only add to the radiation dose delivered to the tissue alongside the primary beam but also, due to their differing momentum directions compared to the primary beam, contribute to dose deposition in normal tissue. Additionally, secondary particles with atomic number (Z) and mass number (A) lower than the primary beam extend the dose contribution beyond the Bragg peak of the target. Consequently, a comprehensive radiation transport analysis is imperative for effective treatment planning. This demands the evaluated data sets of differential and double differential cross sections corresponding to each fragment produced in collision with the isotopes in the tissue, in the therapeutic energy ranges.
\par There are several measurements existing for determining residue cross sections, angular distributions $(\frac{d \sigma}{d \Omega}) $, generating energy spectra $(\frac{d^2 \sigma}{dE_{ej} d\Omega})$ etc. The fragmentation of $^{12}C$ beam with 400 MeV/u energy on an 8 mm thick graphite target was investigated by the FIRST experimental collaboration \citep{ahkummoli, Toppi}  . Fragment identification was achieved through the employment of a $\Delta E$-TOF technique integrated into the FIRST apparatus \citep{FIRST}. The primary beam rejection was executed by using ALADiN magnet. Due to the solid angle acceptance limitations, the experiment could measure angular distribution only up to 6$^\circ$.  Divay et al., \citep{Divay} have measured the differential and double differential cross sections for $^{12}$C-$^{12}$C reactions at 50 MeV/u, using the ECLAN Reaction chamber based facility at GANIL G22 beam line \citep{ECLAN}. Similarly, Dudovet et al., \citep{Dudouet} has measured the angular distribution of fragments at 95 MeV/u energy. There are several other attempts discussing 0$^{o}$ cross secions, total cross sections etc \citep{zero, total, Zeitlin, Webber, LELLIS, Hartmann}. The results from the FIRST experiment reveals a poor mass resolution and hence will lead to ambiguities in the identification of exit channels. The other measurements are also not adequate for identifying the physics mechanism as the information is not enough to fit the Optical Potential formalism. Hence, the current study relied on angular and energy distributions measured from experiments based on ECLAN Reaction chamber. 

\par There have been efforts to reproduce the experimental angular distributions and energy spectrum of fragments, as described in the theoretical work by Dudouet \citep{Dudouet2}, using various models such as QMD \citep{QMD}, BIC \citep{BIC}, and INCL$++$ \citep{INCL}. However, a significant discrepancy remains in the differential and double differential cross sections. The inability to replicate the exact shape indicates an inadequate incorporation of the necessary physics in these models. To address this, a coherently integrated model named QBBC\_ABLA \citep{Ivantchenko2012171,ALLISON2016186}, which includes Fermi breakups also \citep{Fermi}, is attempted here. The first part of this model, Quark-Gluon Based Cascade, accounts for the primary interactions leading to the formation of excited compound nuclei, particle knockout, and the pre-equilibrium. The evaporation of the compound nuclei, generated in the QBBC stage, is estimated using the ABLA Evaporation/Fission Model \citep{ABLA}. Hence, the capability of QBBC\_ABLA model for explaining the fragment production and the kinematic behaviour of it's physics levels has been explored in the present work. 

\section{QBBC\_ABLA Calculations} 
\par The Quantum Chromodynamics-based calculation of $^{12}$C-$^{12}$C reaction has been performed using the QBBC\_ABLA PhysicsList available in Geant4 \citep{Geant4}.  The capability of the QBBC\_ABLA model for predicting fragmentation cross sections and angular distributions has been validated by comparing its results with experimental angular distributions for $^{12}$\text{C}($^{12}$\text{C},xp) and $^{12}$\text{C}($^{12}$\text{C},x$\alpha$).
\begin{figure}
    \includegraphics[width=\columnwidth]{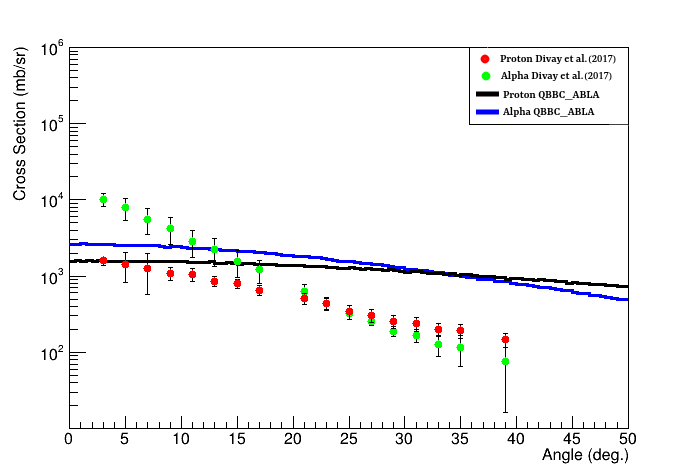}
    \caption{The QBBC\_ABLA calculated proton and $\alpha$ angular distributions of $^{12}$C($^{12}$C,xp) and $^{12}$C($^{12}$C,x$\alpha$) at 50 MeV/u along with measurement by Divay et al.\citep{Divay}}
    \label{fig:ap}
    \end{figure}
     \begin{figure}
    \includegraphics[width=\columnwidth]{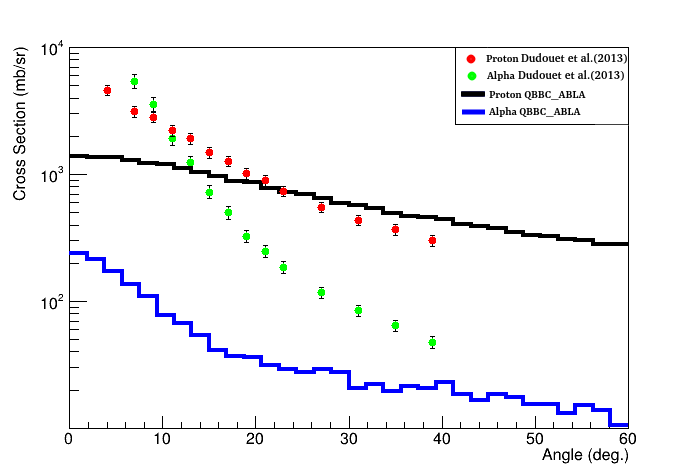}
    \caption{The QBBC\_ABLA calculated proton and $\alpha$ angular distributions of $^{12}$C($^{12}$C,xp) and $^{12}$C($^{12}$C,x$\alpha$) at 95 MeV/u along with measurement by Dudouet et al.\citep{Dudouet}}
    \label{fig:95ap}
\end{figure}

\par The QBBC\_ABLA calculations have been performed by simulating a $^{12}$C beam of 50 and 95 MeV/u fired on a thin graphite target. The first-generation particles were scored for the particle angle, energy, and the particle definition. A total of $10^9$ successive events were simulated for each energy to remove Monte Carlo uncertainities. The fragment events were converted into cross-section units and compared with the available experimental cross sections and particle spectrum retrieved from Divay et al. (EXFOR Entry: \#02408) and Dudouet et al. (EXFOR Entry: \#02170).
\par FIG. \ref{fig:ap} illustrates that the QBBC\_ABLA calculations for proton and $\alpha$ angular distributions along with measurements by Divay et al. This shows the QBBC\_ABLA model calculations, for protons, significantly reproduce the experimental results for forward angles up to 10$^\circ$. However, at higher angles, the theoretical calculations overestimate the results by about an order of magnitude. Similarly, for $\alpha$ angular distributions, the QBBC\_ABLA model underpredicts the cross sections at forward angles below 10$^\circ$ and overpredicts them above 10$^\circ$. In the case of the angular distribution of the 95 MeV/u beam as illustrates in FIG. \ref{fig:95ap}, the proton distribution is reasonably well reproduced at mid-angles but shows significant discrepancies at forward angles. In contrast, the $\alpha$-particle distribution does not achieve agreement with the available experimental measurements.

\begin{figure}
    \includegraphics[width=\columnwidth]{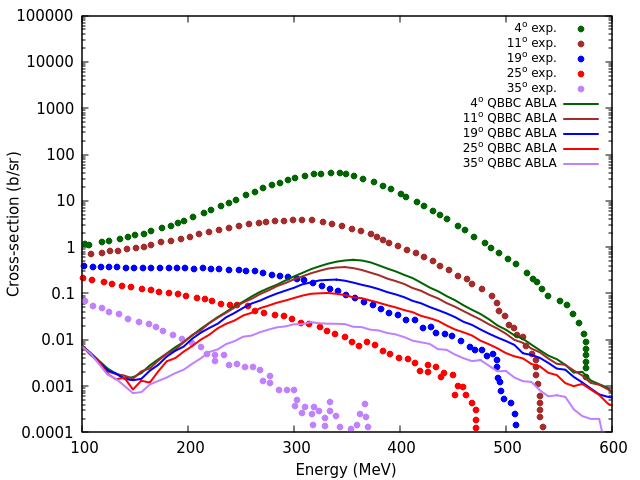}
    \caption{The QBBC\_ABLA calculated $\alpha$ spectrum at different angles for 95 MeV/u $^{12}$C along with the experimental measurements from Dudouet et al., \citep{Dudouet}}
    \label{fig:alpha spectrum different angles}
\end{figure}

\par FIG. \ref{fig:alpha spectrum different angles} compares the $\alpha$ spectra generated using QBBC\_ABLA at various angles with the experimental measurements by Dudouet et al. It can be seen that significant discrepancies have been observed, and these model calculations fail to reproduce the particle spectra across all angles. This failure is further evidenced by the mismatch in angular distributions for energies of 95 MeV/u and mid-backward angles of 50 MeV/u, where the deviations are in the order of several magnitudes. These discrepancies are attributed to the inadequate representation of the underlying physics processes in the reaction mechanism, such as the exclusion of various partial waves and the involvement of breakup modes, particularly the influence of different statistical breakup processes on $\alpha$ production. The QBBC\_ABLA model only holds the Fermi breakups and lacks the sequential breakup mechanisms. The prominence of inelastic and transfer-induced breakup modes can be attributed to the higher interaction energies involved. Within these modes, resonant and continuum breakups might be coupled, requiring the effective determination of the angular distribution and energy spectrum of the emitted $\alpha$ particles. Despite these issues, the ability of this model to reproduce proton angular distributions at lower angles suggests that it partially accounts for pre-equilibrium and knockout processes, although the precise physics model remains incomplete to adequately predict the physics behind how the proton cross sections are evolved. Currently, there is no available data in the literature regarding breakups from different states. The existing experimental measurements primarily focus on angular distributions alone and do not measure particle coincidences to generate relative energy distributions for identifying contributions from different breakup states.
\begin{figure}
    \includegraphics[width=\columnwidth]{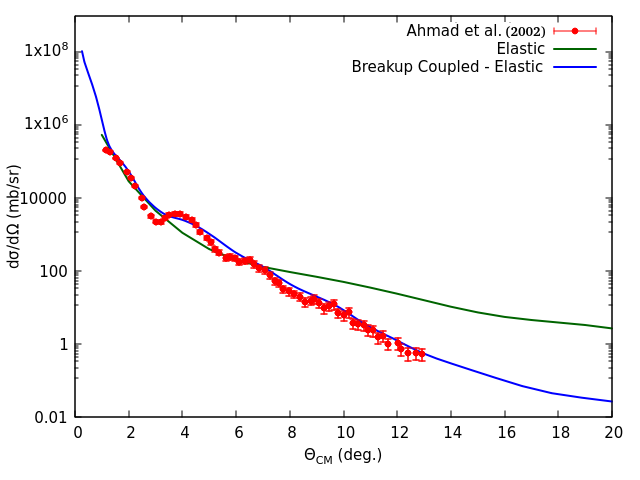}
    \caption{Elastic scattering angular distribution of $^{12}$C-$^{12}$C with and without transfer couplings at 120.75 MeV/u. The experimental data has been adopted from Ahmad et al., \citep{Ahmad}}
    \label{fig:Elastic}
\end{figure}


\section{Alpha Production in Direct Reaction Component}
\par The production of $\alpha$ particles in direct reactions is anticipated to occur through two primary pathways: \textbf{a:} transfer-induced breakup of residual $^8$Be and \textbf{b:} inelastic breakup of $^{12}$C. In the transfer process, an $\alpha$ particle is transferred between the projectile $^{12}$C and the target $^{12}$C, with the interaction coupling through the resonant states of $^{16}$O and $^8$Be. Given that $^8$Be is unbound, the breakup continuum states overlap with the resonant states. For the inelastic breakup, coupling to states above the $^{12}$C $\rightarrow$ $^8$Be + $\alpha$ breakup threshold of 7.37 MeV is considered. Both inelastic and transfer-induced breakups are combined coherently and then coupled to the entrance channel mass partition. The population of individual excited states formed through these mechanisms is calculated using the {\sc fresco} \citep{FRESCO} CDCC-CRC approach. The individual recoil angles of $^8$Be or the forward folding angles of further-decayed $\alpha$ particles were reconstructed in the laboratory frame, and the resulting angular distribution is compared with the adapted experimental data. Before performing the cluster transfer calculations, the elastic angular distributions were reproduced to ensure the capability of {\sc fresco} at these energies, near to the relativistic limits. The experimental data for $^{12}$C$-^{12}$C elastic scattering at 120.75 MeV/u was adapted from Ahmad et al., \citep{Ahmad}. The elastic angular distributions calculated using both the CDCC-CRC method and calculations without transfer coupling are presented alongside the experimental cross section in FIG. \ref{fig:Elastic}. From this figure it is evident that the elastic scattering cross section data is more or less reproduced only when the transfer coupling is incorporated through CDCC-CRC approach.

\subsection{Transfer-Induced Breakup}

\par
\begin{figure}
    \includegraphics[width=\columnwidth]{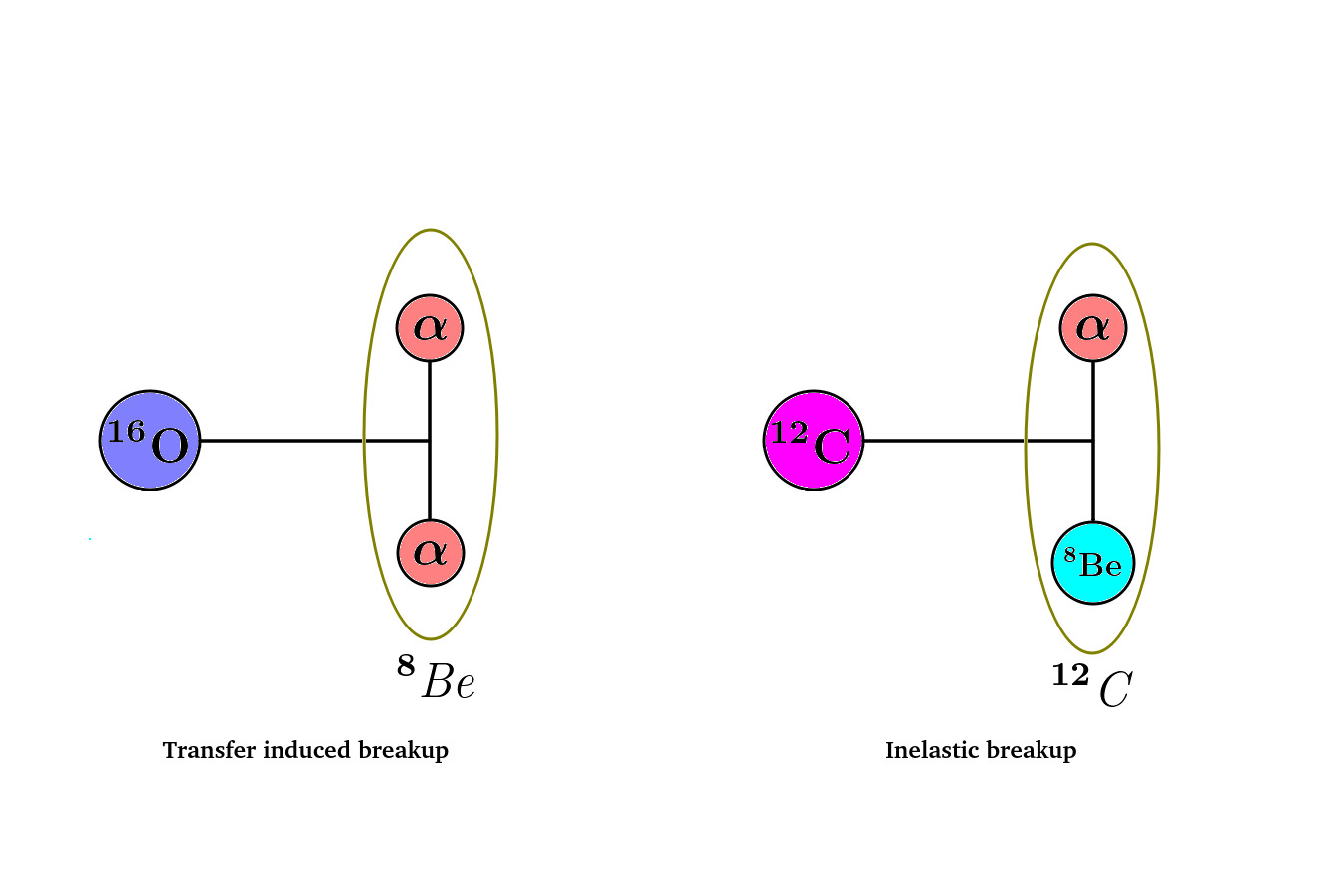}
    \caption{Jacobi coordinate representation of the 3 body interaction in $\braket{\alpha + \alpha - ^{16}O | ^8Be - ^{16}O}$ and $\braket{^8Be + \alpha - ^{12}C | ^{12}C - ^{12}C}$ }
    \label{fig:Jacobi}
\end{figure}

\begin{figure}
    \centering
    \includegraphics[width=\columnwidth]{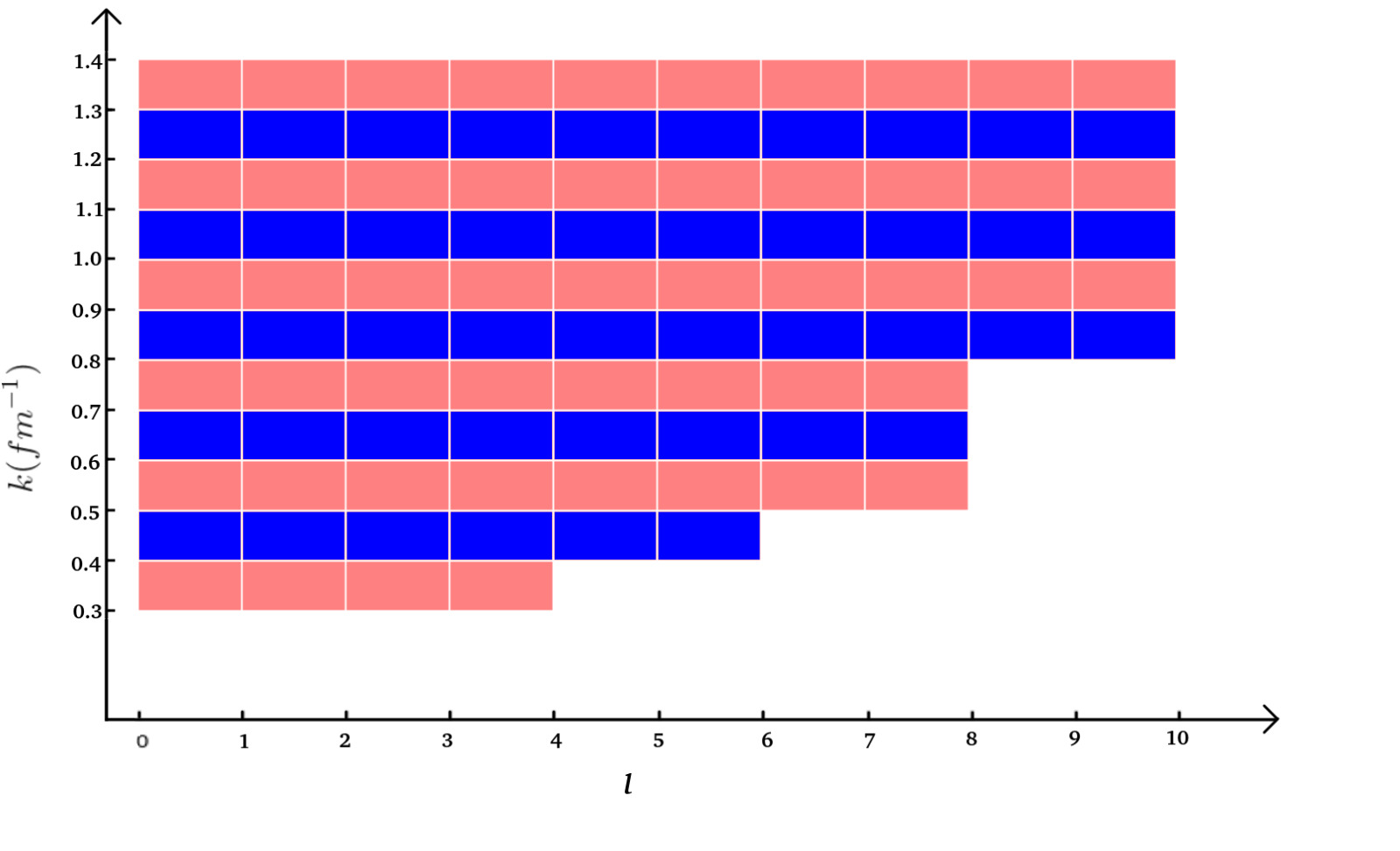}
    \caption{Representation of $\braket{\alpha + \alpha  | ^8Be }$ breakup continuum discretized in 0.1 fm$^{-{1}}$ on k and l}
    \label{fig:cdccaa}
\end{figure}

 A significant pathway for breakup involves the transfer of an $\alpha$ cluster from the projectile to the target (stripping) or vice verca (pickup), leading to the formation of $^8$Be in the exit channel, which can be in its ground or resonant states. This $^8$Be then decays into two $\alpha$ particles through sequential breakups. This transfer-induced breakup mechanism was considered as finite-range transfer due to the higher recoil velocities of the residues. The transferred $\alpha$ particles populate the excited states of $^8$Be and $^{16}$O, with the breakup continuum states generated by the relative movement of $\alpha$-$\alpha$ in $^8$Be also considered. The Jacobi coordinate representation of the three-body interaction for $\braket{\alpha + \alpha - ^{16}O | ^8Be - ^{16}O}$ and $\braket{^8Be + \alpha - ^{12}C | ^{12}C - ^{12}C}$ is illustrated in FIG. \ref{fig:Jacobi}. The discretized states corresponding to $\braket{\alpha + \alpha | ^8Be}$ continuum, binned in radial separation and angular momentum are shown in FIG.  \ref{fig:cdccaa}. Overlaps between the breakup continuum and resonant states are considered, as these overlaps increase the resonant width of CRC states. Calculations are performed using the {\sc fresco} CDCC-CRC approach with an entrance channel mass partition of $^{12}$C+$^{12}$C and an exit channel partition of $^8$Be+$^{16}$O. In the exit channel, $^8$Be is defined as having an $\alpha$ core and $\alpha$ valence, accounting for the breakup continuum. The CRC states include 21 excited states of $^{16}$O up to 13 MeV and 4 excited states of $^8$Be. For a beam energy of 50 MeV/u, states of $^8$Be up to 3.03 MeV and at 95 MeV/u, up to 16.62 MeV, are coupled with $^{16}$O states up to 11.096 MeV. The binning potential for $^{16}$O is defined as p+$^{15}$N.
\par
\begin{figure}
    \includegraphics[width=\columnwidth]{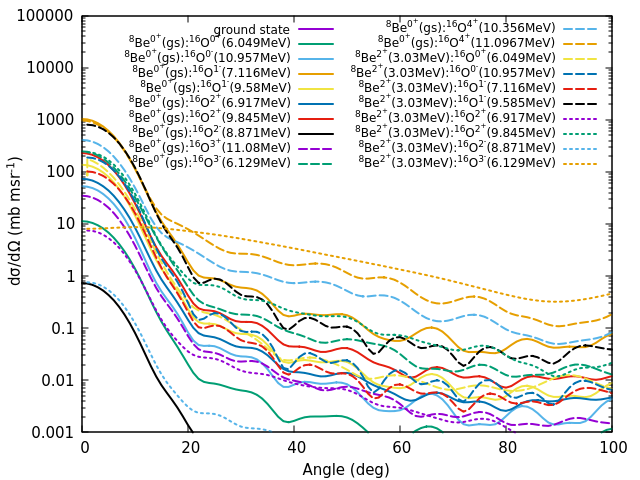}
    \caption{Contribution of different CRC states to the total $\alpha$ cross section, populated through $\alpha$ transfer couplings}
    \label{fig:differential_crc}
\end{figure}

 Contribution of different CRC states to the total $\alpha$ cross section, populated through $\alpha$ transfer couplings and suppressing the population through pre-compound nuclear reaction mechanisms is illustrated in FIG. \ref{fig:differential_crc}.
Each CRC state of the exit channel is coupled with the elastic states of the entrance channel through a spin-transfer coupling. The Optical Model Potential for heavy ions, adapted from Akyuz and Winther \citep{NRV}, is utilized. The breakup continuum up to 20 MeV is discretized based on wave numbers in intervals of 0.1 fm$^{-{1}}$ A spin-spin overlap exists between co-existing resonance and breakup states. The breakup states are defined with Gaussian-shaped potentials for $\alpha$-$\alpha$ interactions. Spectroscopy factors for the $^8$Be-$^{16}$O states are adapted from HFB calculations \citep{HFB}, with an average value of 0.75 used for continuum states. The angle of recoil for $^8$Be is calculated using two-body in-flight decay kinematics, and the $\alpha$ angular distribution is determined.
\par The overlapping factors produced by the two level mixing of single particle states, calculated through HFB Theory has been used as the spectroscopy factors. The quasi particle energies and $J^\pi$ values were matched to the spectroscopic data of corresponding nuclei, for the optimization of HFB code. The spectroscopy factors were further optimized, for fitting the angular distribution, as exclusive cross section data are not available. 

\subsection{Inelastic Breakup}

\par
\begin{figure}
    \centering
    \includegraphics[width=\columnwidth]{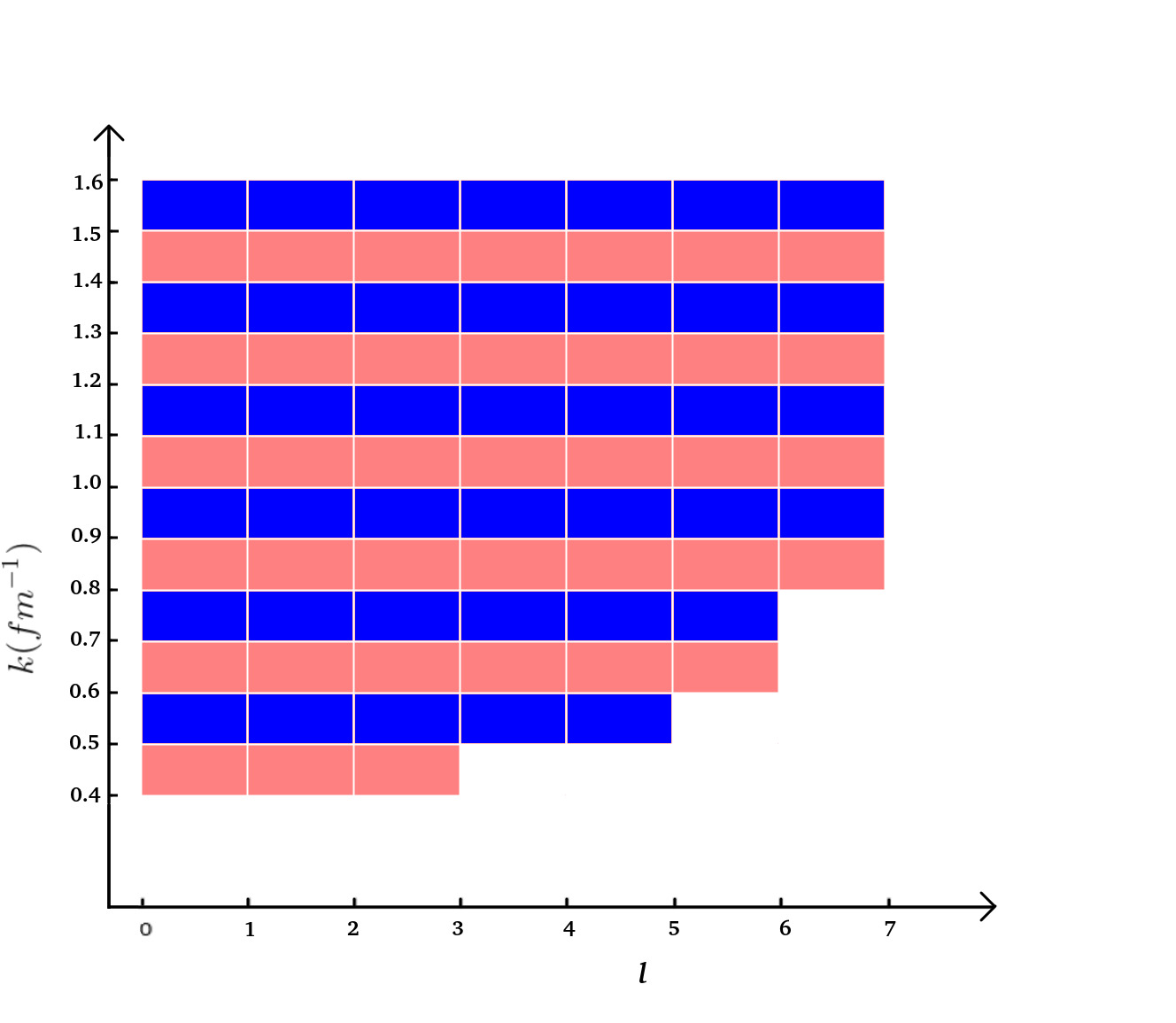}
    \caption{Representation of $\braket{\alpha + ^8Be | ^{12}C }$ breakup continuum discretized in 0.1 fm$^{-{1}}$ on k and l}
    \label{8Bea.jpg}
\end{figure}

 For the inelastic breakup, the elastic-inelastic mass partition is defined as $^{12}$C+$^{12}$C, with the projectile $^{12}$C considered as the $^8$Be-$\alpha$ core-valence structure. The mass partition includes 32 excited states of $^{12}$C including the ground state, defined with a $^{11}$B+$p$ binning potential, to reproduce the radial wave function. The overlap of the $^8$Be+$\alpha$ breakup continuum, above the 7.36 MeV excitation threshold, is also considered. The breakup continuum up to 20 MeV is discretized based on wave numbers in intervals of 0.1 fm$^{-{1}}$ The CDCC states corresponding to the inelastic breakup are illustrated in FIG. \ref{8Bea.jpg}. Each resonant state is coupled to the CDCC states through spin-transfer coupling. For defining the mass partitions, the Optical Model Potential for heavy ions from Akyuz and Winther \citep{NRV} is used here analogous to the transfer-induced breakup. A spin-transfer overlap exists between co-existing resonance and breakup states, and the population of these individual states is calculated using the CDCC-CRC method. Spectroscopy factors for the $^{12}$C states, adapted from the HFB calculations, averaged to 0.75 for continuum states. The three-body folding potential is employed to define discretized continuum states. Above the 7.36 MeV threshold, in-flight three-body decay kinematics is used to calculate $\alpha$ folding angles \citep{Meijer}. The angular distribution of $\alpha$ particles from inelastic breakup is estimated accordingly.

\par At 50 MeV/u, transfer-induced breakup is predominant, while inelastic breakup is negligible. Conversely, inelastic breakup is significant at 95 MeV/u. For 50 MeV/u, the angular distribution of $\alpha$ is reconstructed using in-flight decay kinematics, though it is negligible at 95 MeV/u. Alpha energies are estimated from individual excited states, with cross sections taken from resonance or unbound states. The total angular distribution of $\alpha$ particles has been derived by coherently adding the inelastic and breakup components to the compound nuclear, knockout, and pre-equilibrium contributions, calculated using the QBBC\_ABLA model. The resulting angular distributions and energy spectra are subsequently compared with experimental cross-section data.

\par
\begin{figure}[h]
    \centering
    \includegraphics[width=\columnwidth]{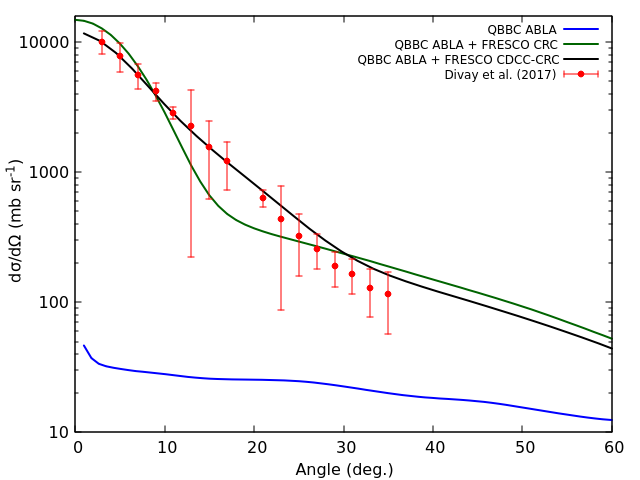}
    \caption{Differential cross-section of $\alpha$ transfer induced breakup at 50 MeV/u $^{12}C$ using {\sc fresco}  CDCC-CRC along with QBBC\_ABLA, compared with Divay et al., \citep{Divay}expeimental values}
\label{fig:50MeV}
\end{figure}

\begin{figure}
    \centering
    \includegraphics[width=\columnwidth]{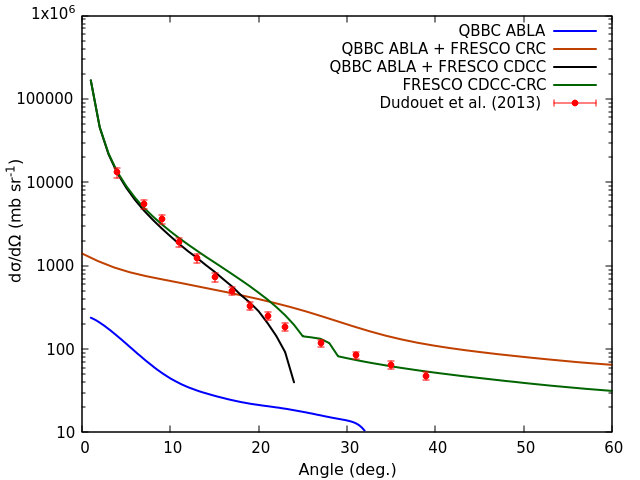}
    \caption{Differential cross-section of inelastic breakup and $\alpha$ transfer induced breakup at 95 MeV/u $^{12}C$ using {\sc fresco}  CDCC-CRC along with QBBC\_ABLA, compared with Dudouet et al., \citep{Dudouet} experimental values}
    \label{fig:95MeV}
\end{figure}

 The incorporation of the $\alpha$ transfer induced breakups and inelastic breakup modes along with the statistical QBBC\_ABLA calculations has sucessfully reproduced the particle spectra and angular distributions measured by Divay et al., \citep{Divay} and Dudouet et al., \citep{Dudouet} Calculation of $\frac{d \sigma}{d\Omega}$, corresponding to 50 MeV/u and 95 MeV/u are illustrated in FIG. \ref{fig:50MeV} and \ref{fig:95MeV}, respectively. This theoretical calculation illustrates that, at 50 MeV/u, the $\alpha$ transfer from the projectile to target, or vice verca, is producing unbound $^8$Be, which enhances the $\alpha$ production. Other than the ground state transfers, the low lying resonant states are also coupled into the transfer mechanism, followed by the breakup. Further, the CDCC states of $^8$Be also couples to the breakup process and act as a spectroscopic overlap to the CRC states. The coupling of low lying states makes the angular distribution highly forward focused. However, the overlapping of CDCC on CRC states makes the angular distribution more broader at mid angles. Corresponding to 95 MeV/u, the inelastic coupling to the breakup states, $\braket{\alpha+^8Be|^{12}C}$, couples to the reaction along with the transfer induced breakup, resulting in the emission of three $\alpha$ particles in the forward direction. The breakup couplings make a pre-compund flux loss, and thereby the compound nucleus formation is inhibited. This effect makes the reduction in the proton and $\alpha$ particle yields at higher angles. This approach accounts for both transfer-induced and inelastic breakup for $\alpha$ production, considering the population of each individual unbound state due to both breakup components, then coherently adding them to obtain the overall $\alpha$ production.
\par
\begin{figure}[h]
    \centering
    \includegraphics[width=\columnwidth]{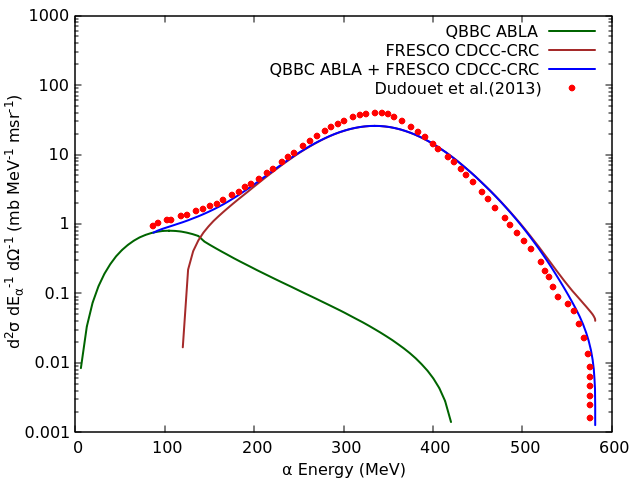}
    \caption{CDCC-CRC + QBBC\_ABLA calculated double differential cross-section ($\alpha$ spectrum) at 4$^0$ showing direct and compound nuclear components, along with experimental measurements from Dudouet et al., \citep{Dudouet}  }
    \label{fig:Aspectrum}
\end{figure}

 The decay of individual continuum and resonant states was accounted for using three-body kinematics, leading to the reproduction of the exit channel alpha spectrum and angular distribution. Furthermore, the double differential cross-section of $\alpha$ at $4^\circ$ for the energy 95 MeV/u is shown in FIG. \ref{fig:Aspectrum}. Here also, the experimental data could be reproduced by incorporating CDCC-CRC calculations along with QBBC\_ABLA, indicating that there is a substantial breakup contribution. At intermediate energies, transfer induced mechanism is the primary driver of the $\alpha$ emission in $^{12}$C-$^{12}$C collisions, while higher energy levels exhibit a balanced interplay between the alpha transfer induced breakup and inelastic breakup. An overlapping of continuum states with the resonant states are also considered to account the higher resonant width of CRC states. These effects collectively contribute to successful matching of the $\alpha$ spectrum with the experimental data from Dudouet et al. The incorporation of spectroscopy factors played a pivotal role in successfully reproducing both the angular distribution as well as the total $\alpha$ spectrum through HFB calculations.
\par
\begin{figure}
    \centering
    \includegraphics[width=\columnwidth]{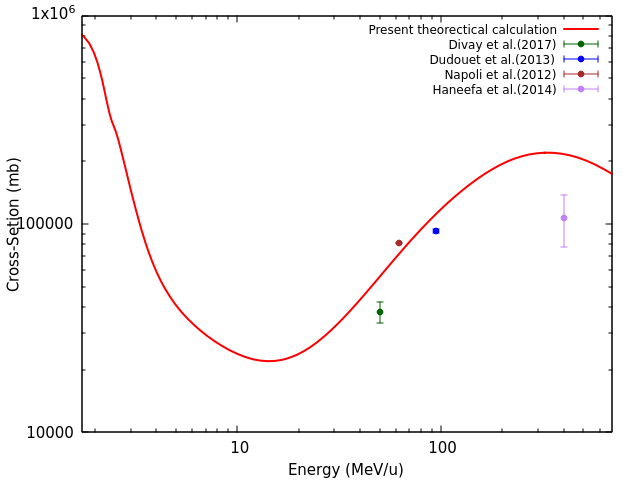}
    \caption{The excitation function for $^{12}C(^{12}C,x\alpha)$ calculated by the coherent addition of the compound nuclear, pre-equilibrium and hadronic component from QBBC\_ABLA and breakup component from {\sc fresco} CDCC-CRC, along with the experimental data from Divay et al., \citep{Divay} for 50 MeV/u, Napoli et al., \citep{Napoli} for 62 MeV/u, Dudouet et al., \citep{Dudouet} for 95 MeV/u and Kummali et al., \citep{ahkummoli} for 400 MeV/u}
    \label{fig:Excitation}
\end{figure}

 Based on the optimized potential parameterization, states, couplings, and spectroscopy factors used to reproduce the angular distributions, the complete excitation function for $^{12}$C($^{12}$C, x$\alpha$) has been calculated. This includes compound, pre-equilibrium, and hadron interaction using the QBBC\_ABLA model, as well as the breakup component from the {\sc fresco} CDCC-CRC. Calculations have been performed for the energy range up to 500 MeV/u and compared with available experimental data taken from \citep{Divay}, \citep{Dudouet}, \citep{ahkummoli} and \citep{Napoli}. The theoretical excitation function, along with available cross sections for the inclusive $\alpha$ production, is illustrated in FIG. \ref{fig:Excitation}.
\par
\begin{figure}
    \centering
    \includegraphics[width=\columnwidth]{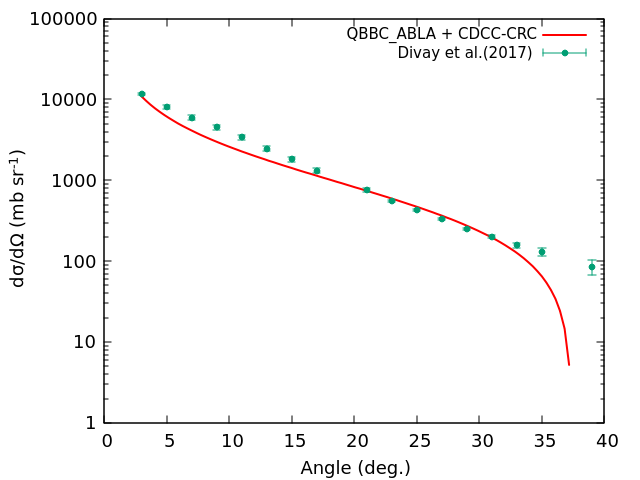}
    \caption{Differential cross-section of alpha production at 50 MeV/u $^{12}C$ beam, on $^{16}O$ using {\sc fresco}  CDCC-CRC along with QBBC\_ABLA then compared with the Divay et al., \citep{Divay} experimental values}
    \label{fig:12C_16O}
\end{figure}

\begin{figure}
    \centering
    \includegraphics[width=\columnwidth]{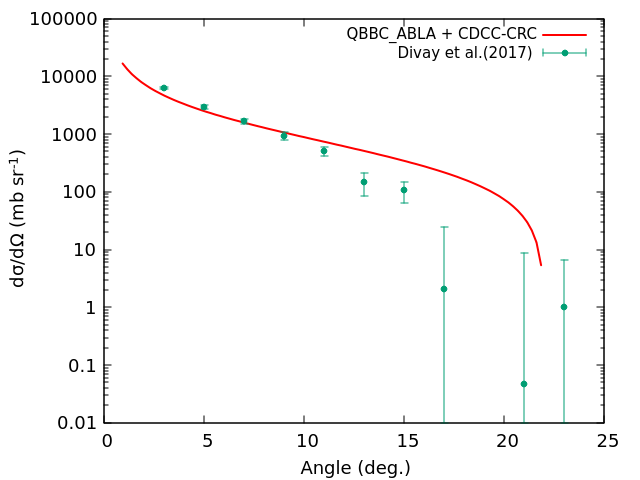}
    \caption{Differential cross section of alpha production at 50 MeV/u $^{12}C$ beam, on $^{1}H$ using {\sc fresco}  CDCC-CRC along with QBBC\_ABLA then compared with the Divay et al., \citep{Divay} experimental values}
    \label{fig:12C_1H}
\end{figure}

  The present approach has been extended to $^{12}$C-$^{16}$O and $^{12}$C-$^{1}$H fragmentation cross sections since it is required to calculate the dose distribution in water and tissue. For holding this CDCC-CRC approach, the inclusive $\alpha$ production from $^{12}$C-$^{16}$O and $^{12}$C-$^{1}$H has also been calculated in the similar manner. In the case of $^{12}$C-$^{16}$O, the $\alpha$ transfer is populating $^{20}$Ne states, along with $^8$Be states. Further the inelastic component also significantly couples to the $\alpha$ production. This approach successfully reproduce the experimental data taken from Divay et al. as shown in FIG. \ref{fig:12C_16O}. The direct $\alpha$ production component in $^{12}\mathrm{C} - ^1\mathrm{H}$ is assumed to be formed through $^{12}\mathrm{C}(p,p')^{12} \mathrm{C} ^* \rightarrow 3 \alpha$, in the inverse kinematics. The population of the unbound $^{12}$C states have been calculated by accounting inelastic couplings and is compared with the experimental data taken from Divay et al. as shown in FIG. \ref{fig:12C_1H}. The  $\alpha$ cross sections are successfully reproduced at the forward angles but a mismatch exists at the backward angles due to the high uncertainities expected in the experimental values. The potential parameters were adapted from the evaluation, and the spectroscopy factors from the HFB calculations. 


\begin{figure}
    \centering
    \includegraphics[width=\columnwidth]{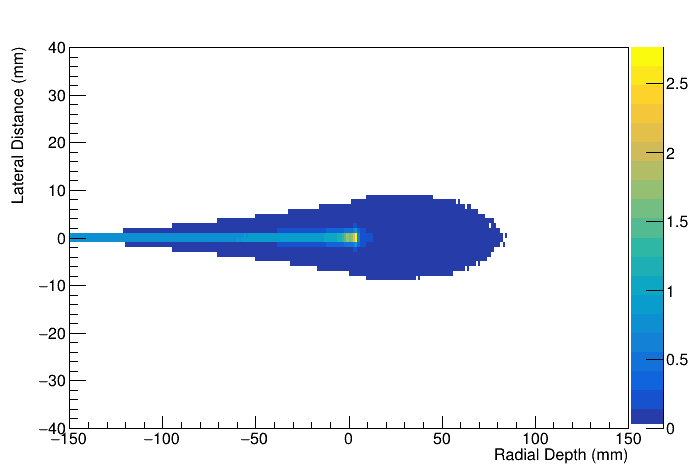}
    \caption{The z-x planar dose deposition of $^{12}C$ on water }
    \label{Total_tissue_w_enhancement.png}
\end{figure}
\begin{figure}
    \centering
    \includegraphics[width=\columnwidth]{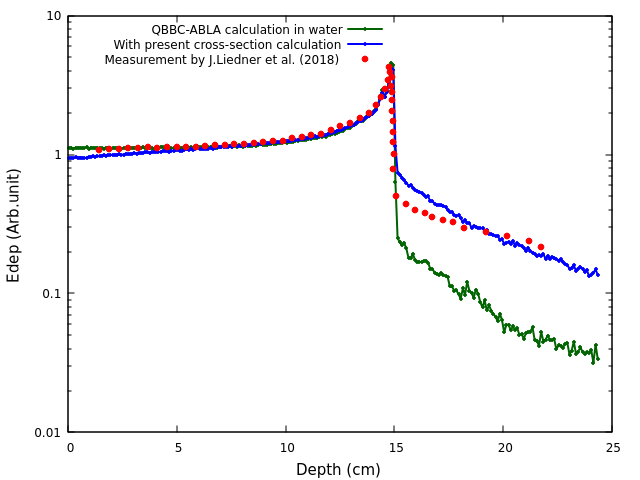}
    \caption{The comparison of radial energy deposition of $^{12}C$ on water along with QBBC\_ABLA calculations and sequential breakup. The experimental data has been adopted from Liedner et al.\citep{Leidner_2018}}
    \label{water_braggpeak_comparison.png}
\end{figure}
\begin{figure}
    \centering
    \includegraphics[width=\columnwidth]{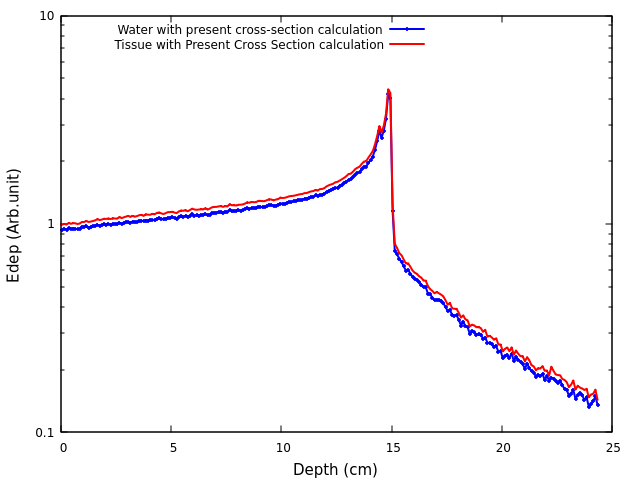}
    \caption{Radial energy deposition in water and tissue}
    \label{Water_tissue_comparison.png}
\end{figure}
 \section{Dose Calculations}
\par Given that the experimental measurements are only available in water, simulations of $^{12}$C on water at 280 MeV/u were conducted by incorporating the breakup components as an additional physics process alongside the standard PhysicsList in Geant4. The CDCC-CRC calculated cross sections, for the new physics process is given as a numerical input to Geant4. The typical geometry of 30 cm $\times$30 cm $\times$30 cm has been considered as the target, in tracking mode. The cut values of 0.1 mm, optimum compared to the mean free paths of primary, secondary and the $\delta -$rays has been chosen for the simulations. The energy deposit, position vertex, particle id, and the kinetic energy stamp of the interacting particle are scored in each step, and pushed to the root file. A two dimensional histogram between radial depth and X-position, gated to unity Y-position, weighted with the energy deposit has been constructed and is shown in FIG. \ref{Total_tissue_w_enhancement.png}. The energy deposit in each voxal were divided with the mass of the voxal for obtaining the absolute dose deposit. An appropriate 2D gate, considering the beam lateral blow up has been defined and the gated region has been projected on to the radial axis, for estimating the radial dose deposition profile. The obtained radial dose profile has been compared with the measurement by Liedner et al., \citep{Leidner_2018} as shown in FIG.  \ref{water_braggpeak_comparison.png}. Since the available experimental data by Liedner et al. is based on PeakFinder and published as an arbitrary value, the current simulation result has been normalized to the value at the Bragg peak, and the dose profile has been compared. In order to identify the enhancement due to breakup coupling, the standard QBBC\_ABLA based simulation is also performed with the default parameters, and presented in the FIG. \ref{water_braggpeak_comparison.png}. Similar kind of simulation has been performed for tissue, adapting original elemental composition of tissue, other than considering the equivalent media in geometry and material definition. This is further compared with the breakup incorporated energy deposition curve simulated on water as given in FIG.  \ref{Water_tissue_comparison.png}. Here the breakup couplings associated to the $^{12}C-^{16}O$ reaction is also incorporated to the physics process. The individual fragmentation components, associated to the dose deposition has been identified through particle ids acquired. The lateral components of the dose deposition at beam entrance, Bragg peak region and dose tail region are also analysed.
\par The present simulation accounting breakup components, gives a better correspondence to the experimentally measured energy deposit curve in water medium, especially at the initial depths. The dose tail region of the energy deposit curve produced by the fragments, has been well reproduced. Since the $\alpha$ fragments, which are produced before the Bragg peak region, being lighter than the primary $^{12}$C beam, they are extending beyond the Bragg peak point. Further, it also depends upon the energy spectrum produced by the fragmentation process, in all phases of its interaction, and leads to the propagation of these into the tail region.
\par In the present calculation, most of the tail region has been reproduced. However, the immediate tail after the Bragg peak has been slightly exaggerated. This has been anticipated because of the production of other lighter nuclei such as $^7$Li, $^3$He etc., through the direct reaction mechanism. Here, the production of these lighter nuclei are accounted only through compound nuclear and Fermi breakup mechanisms excluding the quantum mechanical approach.  The involvement of these elements are absent in the present CDCC-CRC calculation, due to these ambiguities in the production mechanism, unavailability of potentials and spectroscopy factors etc. The involvement of many-body problem, beyond 3 body is also a reason for excluding.
\par In summary, this study successfully reproduced the dose tail region by accurately accounting for the physics of fragmentation. The production of $\alpha$ fragments, which occur in significant quantities due to breakup coupling, is a key factor. Transfer-induced breakups and inelastic breakups substantially enhance $\alpha$ production. Given their higher initial velocity, these $\alpha$ fragments are forward-focused, resulting in a significant dose tailing beyond the Bragg peak. The angular distribution and energy spectrum have been theoretically reproduced, and the excitation function for inclusive $\alpha$ production has been estimated. This theoretical framework has been utilized for simulating dose deposition, achieving a higher level of reproducibility.

\par The present study suggests that, in addition to the phenomenological and statistical PhysicsLists, precise physics in the domain of hadron interactions, breakup modes, compound nucleus formation, and pre-equilibrium processes should also be integrated. This must be thoroughly characterized through exclusive experimental measurements for each breakup mode from each resonant and continuum states. The measured exclusive cross sections should be reproduced using CDCC-CRC calculations within a many-body approach for multi-fragmentation. PhysicsLists for simulations should be grounded on these physics-driven approaches rather than purely phenomenological methods, whether for total Monte Carlo simulations or the Multi-Group Kernel approach.

\section*{Acknowledgements}
Authors acknowledge the support of HPC group for providing the high performance computing facility.
\par Authors express gratitude to Dr. Sanjib Muhuri, High Energy Physics Group, Variable Energy Cyclotron Centre, Kolkata, India for the initial discussions on incorporating the transfer coupled cross sections to Geant4 PhysicsLists.
\par Authors express thanks to Dr. N. Otuka, Nuclear Data Section, IAEA for the support given on identifying cross section data sets and their EXFOR entries. 
\bibliographystyle{apsrev4-2}
\bibliography{main}
\end{document}